\documentclass[12pt]{JHEP3}
\usepackage{amsmath}
\usepackage{graphicx}

\newcommand{\rbvar}{\rho}

\title{Nonlinear viscous hydrodynamics in various dimensions using AdS/CFT}

\author{Michael Haack and Amos Yarom \\
	Ludwig-Maximilians-Universit\"at, Department f\"ur Physik, Theresienstrasse 37, 80333 M\"unchen, Germany \\
	E-mail: \email{michael.haack@physik.uni-muenchen.de, amos.yarom@physik.uni-muenchen.de}}

\abstract{We compute coefficients of two-derivative terms 
in the hydrodynamic energy momentum tensor of a viscous fluid which has an AdS${}_D$ dual 
with $3 \leq D \leq 7$. For the case of $D=3$ we obtain an exact AdS${}_3$ black hole solution, valid to all orders in a derivative expansion, dual to a perfect fluid in $1+1$ dimensions.}


\preprint{LMU-ASC 39/08}

\begin{document}

\section{Introduction and summary}
The gauge-string correspondence, or AdS/CFT, has proved to be an effective tool for studying strongly coupled dynamics of gauge-theories. 
In particular, conformal gauge theories at finite temperature and long wavelengths 
have a hydrodynamic description which can be captured by AdS black holes.
An important early application of such a dual description can be found in \cite{Policastro:2001yc}, where the authors computed the shear viscosity to entropy ratio of the $\mathcal{N}=4$ supersymmetric Yang-Mills plasma at large 't Hooft coupling,
\begin{equation}
\label{E:StoE}
	\frac{\eta}{s} = \frac{1}{4\pi}\ .
\end{equation}
This ratio is quite robust, retaining its (strong coupling) value of $1/(4\pi)$ in many theories which admit a holographic dual, in any dimension, \cite{Buchel:2003tz,Kovtun:2003wp,Buchel:2004qq,Benincasa:2006fu,Son:2006em}. 
More recently, the authors of \cite{Baier:2007ix, Bhattacharyya:2007jc, Natsuume:2007ty,Natsuume:2008iy,Loganayagam:2008is, VanRaamsdonk:2008fp,Bhattacharyya:2008ji} 
calculated some of the higher order hydrodynamic coefficients in gauge theories with AdS${}_D$ duals.
It is natural to inquire how the higher order coefficients depend on the dimensionality of spacetime. The $D=5$ and $D=4$ cases have already been addressed in \cite{Bhattacharyya:2007jc} and \cite{VanRaamsdonk:2008fp}, respectively, and in \cite{Natsuume:2007ty} one of the four second order coefficients was calculated for $D=4,5$ and $7$.
The purpose of this work is to complete the picture and see how the second order coefficients depend on the spacetime dimension of the boundary theory.

Our method of computation is an immediate extension of the work in \cite{Bhattacharyya:2007jc}. There, an algorithm was presented to  construct asymptotically AdS${}_5$ black holes which are dual to a non-trivial flow of a viscous fluid in four dimensional conformal gauge theories.\footnote{
In the following we use the term black hole to describe a geometry with an event horizon. Thus, we do not distinguish between black holes and black branes; the event horizon of the black holes we consider is flat.
} 
The algorithm involves (perturbatively)  extending the known asymptotically AdS${}_5$ boosted black hole solutions by allowing the temperature and boost parameters to vary slowly over the spacetime coordinates transverse to the radial AdS direction. 
A detailed analysis of the geometric properties of these solutions can be found in \cite{Bhattacharyya:2008xc}.
In \cite{VanRaamsdonk:2008fp} these new black hole solutions were constructed for $D=4$ and in this work we show that such a construction is possible in $3 \leq D \leq 7$. The $D=3$ case is special---as expected on general grounds \cite{Banados:1992gq,Carlip:1995qv} the black hole solution we find is diffeomorphic to the AdS${}_3$ static black hole. 
In section \ref{S:AdS3} we discuss this in more detail.

Explicit string/M-theory realizations of conformal field theories dual to supergravity theories with an empty AdS${}_D$ solution  are available for $D=3,4,5$ and $7$ (in the appropriate limit).
Concrete dualities in $D=6$ are less straightforward to obtain. For instance, the AdS/CFT correspondence based on $D4$-branes leads to a non-trivial dilaton background. Perhaps the more general ideas put forward in \cite{Polyakov:1997tj,Polyakov:1998ju} could be used to obtain concrete examples. 
However, as stressed in \cite{VanRaamsdonk:2008fp}, the long wavelength limit of the new black hole solutions is governed by conformal fluid dynamics 
(with particular higher order transport coefficients) 
even without an explicit realization of AdS${}_D$/CFT${}_{D-1}$. This is very reminiscent of the membrane paradigm \cite{Price:1986yy,Thorne:1986iy,Parikh:1997ma} discussed in a similar context in \cite{Kovtun:2003wp,Starinets:2008fb}.

As in \cite{Bhattacharyya:2007jc}, we use the Eddington-Finkelstein coordinate system to construct the black hole solutions. With it, it is easier to ensure that the metric is smooth everywhere.
Since it is standard practice to use the Fefferman-Graham coordinate system when using the prescription developed in \cite{Gubser:1998bc,Witten:1998qj,deHaro:2000xn,Bianchi:2001de,Bianchi:2001kw} to compute boundary observables, we go over the bulk-to-boundary mapping in the Eddington-Finkelstein coordinate system in section \ref{S:BtB}. This analysis will also allow us to simplify the calculations in later sections.

In section \ref{S:Main} we derive our main results which are the second order hydrodynamic coefficients of the energy momentum tensor of a conformal fluid dual to an AdS${}_D$ black hole. These can be parameterized as follows.
Consider the most general energy momentum tensor of a conformal fluid. We denote the temperature of the fluid by $T$, and its velocity field by $u^{\mu}$ with $u^{\mu}u_{\mu}=-1$. Greek indices run from $\mu = 0,\ldots,d-1 = D-2$ and are raised and lowered with the Minkowski metric. In the static case, when the temperature and velocity field are constant, tracelessness implies that the energy momentum tensor must take the form
\begin{equation}
\label{E:Perfectfluid}
	\langle T_{\mu\nu} \rangle
	=
	p_0 T^{d} \left(d u_{\mu}u_{\nu}+\eta_{\mu\nu}\right)\ ,
\end{equation}
where $p_0$ is a dimensionless constant. In the rest frame of the fluid the energy density is
\begin{equation}
	e_0 T^d = (d-1) p_0 T^d.
\end{equation}
In the non-static case, where the temperature and velocity field are allowed to vary, $u^{\mu}=u^{\mu}(x^{\nu})$, $T=T(x^{\nu})$, one can expand the energy momentum tensor of the fluid in derivatives of the velocity field and of the temperature. A classification of all possible terms up to two orders in the derivatives has been carried out in \cite{Baier:2007ix}: in the Landau frame where
\begin{equation}
\label{E:Landau}
	u^{\mu} \langle T_{\mu\nu} \rangle = -p_0 T^d u_{\nu}\ ,
\end{equation}
and for a conformally flat background, this energy momentum tensor takes the form
\begin{equation}
\label{E:Viscousfluid}
	\langle T_{\mu\nu} \rangle
	=
	p_0 T^{d} \left(d u_{\mu}u_{\nu}+\eta_{\mu\nu}\right)
	-\eta\, \sigma_{\mu\nu}
	+\eta \tau_{\Pi}\, \Sigma^{(0)}_{\mu\nu}
	+\lambda_1\, \Sigma^{(1)}_{\mu\nu}
	+\lambda_2\, \Sigma^{(2)}_{\mu\nu}
	+\lambda_3\, \Sigma^{(3)}_{\mu\nu}\ ,
\end{equation}
where
\begin{equation}
 	\sigma_{\mu\nu} = 2\partial_{\langle \mu}u_{\nu\rangle}\ , \quad
	\omega_{\mu\nu} = \frac{1}{2}P_{\mu}^{\lambda}P_{\nu}^{\sigma}\left(\partial_{\lambda}u_{\sigma}-\partial_{\sigma}u_{\lambda}\right)
\end{equation}
and
\begin{align}
\notag
\label{E:Secondorderterms}
    \Sigma^{(0)}_{\mu\nu} &= {}_{\langle}u^{\lambda}\partial_{\lambda} \sigma_{\mu\nu\rangle}+\frac{1}{d-1}\sigma_{\mu\nu}\partial_{\lambda}u^{\lambda} \ , \\
\Sigma^{(1)}_{\mu\nu} &=\sigma_{\langle \mu \lambda}\sigma^{\lambda}_{\nu\rangle}\ ,\quad
    \Sigma^{(2)}_{\mu\nu} =\sigma_{\langle \mu \lambda}\omega^{\lambda}_{\nu\rangle}\ ,\quad
    \Sigma^{(3)}_{\mu\nu} =\omega_{\langle \mu \lambda}\omega^{\lambda}_{\nu\rangle}\ .
\end{align}
Here $P_{\mu\nu}$ is the projection operator
\begin{equation}
	P_{\mu\nu}=u_{\mu}u_{\nu}+\eta_{\mu\nu}\ ,
\end{equation}
and angular brackets denote a traceless projection onto the space orthogonal to $u_{\mu}$ so that
\begin{equation}
    A_{\langle \mu \nu \rangle}
    = P_{\mu}^{\lambda}P_{\nu}^{\sigma}\frac{1}{2}\left(A_{\lambda\sigma}+A_{\sigma\lambda}\right)
      -\frac{1}{d-1}P_{\mu\nu}P^{\lambda\sigma}A_{\lambda\sigma}
\end{equation}
satisfies $\eta^{\mu\nu} A_{\langle \mu \nu \rangle} = 0$ and $u^{\mu}A_{\langle \mu\nu \rangle} = 0$. Our notation closely follows that of \cite{Baier:2007ix} and differs from that of \cite{Bhattacharyya:2007jc,VanRaamsdonk:2008fp} by a factor of two in the definition of $\sigma_{\mu\nu}$.
The astute reader might worry that none of the terms in \eqref{E:Secondorderterms} contain derivatives of the temperature. This is because derivatives of the temperature may be exchanged with derivatives of the velocity field once the stress-energy tensor is conserved. See \cite{Baier:2007ix} for details.

Thus, according to \cite{Baier:2007ix} the energy momentum tensor of a viscous fluid \eqref{E:Viscousfluid} expanded to second order in derivatives of the velocity and temperature fields, is parameterized by five coefficients $\eta$, $\tau_{\Pi}$ and $\lambda_i$ with $i=1,\ldots,3$.
For the fluids considered in this paper, we find
\begin{subequations}
\label{E:Results}
\begin{align}
	\frac{\eta}{s} &= \frac{1}{4\pi}\\
	\frac{\eta\tau_{\Pi}}{c} &= \frac{d}{16\pi^2(d-1)^2}+
		\frac{d}{32\pi^2(d-1)^2}
		\begin{cases}
		\frac{\pi}{3\sqrt{3}}-\ln 3 & \qquad d=3\\
		-\ln 2 & \qquad d=4 \\
		-\frac{\pi}{5}\sqrt{1-\frac{2}{\sqrt{5}}}+\frac{\coth^{-1}\sqrt{5}}{\sqrt{5}}-\frac{1}{2}\ln 5 & \qquad d=5\\
		-\frac{\pi}{6\sqrt{3}}-\frac{1}{2}\ln 3 & \qquad d=6
		\end{cases}\\
	\frac{\lambda_1}{c} &= \frac{d}{32\pi^2(d-1)^2}
		\phantom{\frac{1}{\mbox{\Huge{A}}}}\\
\label{E:lambda2}
	\frac{\lambda_2}{c} &= 2 \left( \frac{\eta \tau_{\Pi}}{c}-\frac{d}{16\pi^2(d-1)^2} \right) \\
\label{E:lambda3}
	\lambda_3 & = 0
\end{align}
\end{subequations}
with $s$ the entropy density and $c$ the second derivative of the energy density:
\begin{equation}
\label{E:sdensity}
	s= \frac{\partial}{\partial T} (p_0 T^d)\ ,\quad
	c= \frac{\partial^2}{\partial T^2} (e_0 T^d)\ .
\end{equation}
Our choice for normalizing the second order coefficients by $c$ is arbitrary: we could have also normalized the coefficients by the derivative of the entropy density which would have resulted in different factors of $d-1$. 
Our value for $\eta$ agrees with \cite{Policastro:2001yc} and is not new. The values for $\tau_{\Pi}$ and $\lambda_i$ for $d=3$ and $d=4$ have been computed in \cite{VanRaamsdonk:2008fp,Bhattacharyya:2007jc,Baier:2007ix}, and $\tau_{\Pi}$ has been computed for $d=3,\,4$ and $6$ in \cite{Natsuume:2007ty}. Our results are in complete agreement with those values.

Note that in $d=3$ one has $\Sigma^{(1)}=\Sigma^{(3)}=0$ identically, and in $d=2$ one has $\sigma=\Sigma^{(0)}=\Sigma^{(i)}=0$. In fact, in a $1+1$ dimensional conformal theory there can be no transport coefficients since there is no non-trivial two dimensional transverse, traceless, symmetric  matrix. A similar argument based on finite temperature correlators can be found in \cite{Herzog:2007ij}. 
Thus, for $d=2$, we expect no higher derivative corrections to the energy momentum tensor \eqref{E:Perfectfluid} and indeed, in section \ref{S:AdS3} we find exact black hole solutions to the Einstein equations valid to all 
 orders in a derivative expansion. 

In \cite{Baier:2007ix} a weak coupling analysis of second derivative terms was carried out using kinetic theory. By using a multipole expansion of the Boltzmann equation, it was shown that a conformal theory in $3+1$ dimensions should satisfy 
\begin{equation} 
\label{E:kinetic}
	\lambda_3 = 0\ ,\quad
	\lambda_2 = -2 \eta \tau_{\Pi}\ .
\end{equation}
According to \eqref{E:lambda3} the first of these equalities is satisfied also at strong coupling in any dimension. 
The weak coupling result for $\lambda_2$ differs from the strong coupling result, though we note that in both cases the sign of $\tau_{\Pi}$ is opposite to that of $\lambda_2$.


\section{The bulk to boundary correspondence}
\label{S:BtB}
The algorithm introduced in \cite{Bhattacharyya:2007jc} for finding the black hole solutions dual to fluid dynamics is very similar in spirit to the derivative expansion of the boundary energy momentum tensor described above.
Consider a Poincar\'e patch of a boosted AdS${}_D$ black hole in Eddington-Finkelstein coordinates
\begin{equation}
\label{E:StaticBH}
	ds^2 = -r^2 \left(1-\frac{1}{r^{D-1}b^{D-1}}\right)u_{\mu}u_{\nu}dx^{\mu}dx^{\nu}
	       +r^2 P_{\mu\nu} dx^{\mu}dx^{\nu}
	       - 2 u_{\mu}dx^{\mu}dr
\end{equation}
with
\begin{equation}
	P_{\mu\nu}=u_{\mu}u_{\nu}+\eta_{\mu\nu}\ , \quad u_{\mu}u^{\mu}=-1\ ,
\end{equation}
$3 \leq D \leq 7$ and constant $u_{\mu}$ and $b$. Recall that $\mu$ runs from $0$ to $D-2$.
The Hawking temperature is related to the parameter $b$ by
\begin{equation}
\label{E:btoT}
	b = \frac{D-1}{4\pi T}\ .
\end{equation}
The metric in \eqref{E:StaticBH} is a solution to the Einstein equations
\begin{equation}
\label{E:Einstein}
	R_{MN}+(D-1)g_{MN}=0\ ,\quad R=-D(D-1)
\end{equation}
in the absence of matter.
Roman indices take the values $M=0,\ldots,D-1$.
As explained in \cite{Bhattacharyya:2007jc}, the Eddington-Finkelstein coordinate system is more useful in this case than the standard Fefferman-Graham coordinate system since apart from the curvature singularity at $r=0$, the metric is finite everywhere and the determinant is non-vanishing; if we consider small perturbations of this metric then, as long as the perturbations are finite everywhere, we are assured that only $r=0$ will be a curvature singularity.

The boundary theory stress tensor dual to this solution can be obtained by the prescription of \cite{Balasubramanian:1999re,deHaro:2000xn} which is based on \cite{Gubser:1998bc,Witten:1998qj}. The result, written in covariant form, is
\begin{equation}
\label{E:KmnToTmn}
	\langle T_{\mu\nu} \rangle = \lim_{r \to \infty} \left[ \frac{r^{(D-3)}}{\kappa_D^2} \left(K_{\mu\nu}-K\gamma_{\mu\nu}-(D-2)\gamma_{\mu\nu}\right)\right] ,
\end{equation}
where $\kappa_D$ is the $D$ dimensional gravitational coupling constant ($\kappa_D^2 = 8 \pi G_D$),
\begin{equation}
\label{E:extcurve}
K_{\mu\nu} = -\frac{1}{2n} (\partial_r \gamma_{\mu\nu}
- \nabla_\mu n_\nu
-\nabla_\nu n_\mu)
\end{equation}
are the $\mu$,$\nu$ components of the extrinsic curvature of a hypersurface close to the AdS boundary, $K=K_{\mu\nu}\gamma^{\mu\nu}$ 
and $\gamma_{\mu\nu}$, $n$ and $n_\mu$ are the
boundary metric, lapse function and shift functions, respectively. 
In our setup they are defined by writing the bulk metric as
\begin{equation}
\label{E:ADM}
ds^2 = n^2 dr^2 + \gamma_{\mu \nu}(dx^\mu + n^\mu dr)
(dx^\nu + n^\nu dr)\ .
\end{equation}
See \cite{Brown:1992br} for details.
Under the equations of motion the energy momentum tensor will be traceless \cite{Balasubramanian:1999re,deHaro:2000xn}, which implies that
\begin{equation}
\label{E:Traceless}
	K = -(D-1)\ .
\end{equation}
Using \eqref{E:KmnToTmn} and \eqref{E:Traceless}, the energy momentum tensor dual to the black hole \eqref{E:StaticBH} reads
\begin{equation}
\label{E:TmnPF}
	\langle T_{\mu\nu} \rangle  = \frac{1}{2\kappa_D^{2}} \frac{1}{b^{d}}\left(d\, u_{\mu}u_{\nu}+\eta_{\mu\nu}\right).
\end{equation}
At this point, by comparing \eqref{E:Perfectfluid} to \eqref{E:TmnPF} and using the relation
\eqref{E:btoT}, we find that
\begin{align}
\label{E:p0c}
	p_0 & = \frac{1}{2\kappa_D^2}\left(\frac{4\pi}{d}\right)^d \nonumber \\
	    & = \frac{\Gamma(d/2)^3}{4\pi^{d/2}\Gamma(d)}\frac{(d-1)}{d(d+1)}
		\left(\frac{4\pi}{d}\right)^d \mathfrak{c}\ ,
\end{align}
where $\mathfrak{c}$ is the central charge of the conformal theory and in the last line we have used the results of \cite{Kovtun:2008kw}. Thus, the black holes \eqref{E:StaticBH} are dual to fluids with a constant velocity field and temperature.

To obtain gradient corrections to \eqref{E:TmnPF} we need to allow the inverse temperature $b$ and the velocity field $u^{\mu}$ to vary in space and time. Of course, once we do that \eqref{E:StaticBH} will no longer be a solution to the Einstein equations \eqref{E:Einstein}. The algorithm proposed in \cite{Bhattacharyya:2007jc} is to expand the metric of the black hole in derivatives of the velocity field and temperature, and solve the Einstein equations \eqref{E:Einstein} order by order in a derivative expansion.
To parameterize the corrections to \eqref{E:StaticBH} once the gradients are non vanishing, we first fix a gauge where $g_{rr}=0$ and $g_{r\mu} \propto u_{\mu}$. Then the most general line element that can be obtained, takes the form
\begin{multline}
\label{E:Corrected}
	ds^2 = r^2 k[r,u^{\lambda},b] u_{\mu}u_{\nu}dx^{\mu}dx^{\nu}
	       +r^2 h[r,u^{\lambda},b] P_{\mu\nu} dx^{\mu}dx^{\nu}
	       +r^2 \pi_{\mu\nu}[r,u^{\lambda},b] dx^{\mu}dx^{\nu} \\
           +r^2 j_{\sigma}[r,u^{\lambda},b]\left(P^{\sigma}_{\mu}u_{\nu}+P^{\sigma}_{\nu}u_{\mu}\right)dx^{\mu}dx^{\nu}
           -2S[r,u^{\lambda},b] u_{\mu}dx^{\mu}dr\\
	   \equiv
	   r^2 g_{\mu\nu}dx^{\mu}dx^{\nu}-2S[r,u^{\lambda},b] u_{\mu}dx^{\mu}dr\ .
\end{multline}
The functions $S$, $k$, $h$, $j_{\mu}$ and $\pi_{\mu\nu}$ are determined order by order in a gradient expansion of the velocity field and the temperature. The tensor $\pi_{\mu\nu}$ is symmetric, traceless and transverse to the velocity field, $u^{\mu}\pi_{\mu\nu}=0$. We shall use a superscript $(n)$ to denote the $n$'th order term in a (boundary coordinate) derivative expansion of the various fields $S$, $k$, $h$, $j_{\mu}$ and $\pi_{\mu\nu}$. Thus, according to \eqref{E:StaticBH} we have
\begin{align}
\label{E:zeroorder}
    k^{(0)}[r,u^{\lambda},b] &= -\left(1-\frac{1}{r^{D-1}b^{D-1}}\right)\ ,&
    S^{(0)}[r,u^{\lambda},b] &= 1\ ,&
    h^{(0)}[r,u^{\lambda},b] &= 1\ ,\nonumber \\
    j^{(0)}[r,u^{\lambda},b] &= 0\ ,&
    \pi^{(0)}[r,u^{\lambda},b] &= 0\ .
\end{align}
Note that in \eqref{E:Corrected} there is still some gauge freedom in reparameterizing the radial coordinate. We shall fix this gauge freedom later.

Once $S$, $k$, $h$, $j_{\mu}$ and $\pi_{\mu\nu}$ are known we can determine the higher order corrections to the boundary theory stress tensor via \eqref{E:KmnToTmn} and \eqref{E:Traceless}. Before proceeding, we take the time to simplify our expression \eqref{E:KmnToTmn} for the energy momentum tensor. Let us start by expanding the extrinsic curvature and the metric in a series expansion near the boundary (large $r$), denoting the coefficient of the $1/r^{m}$ term with a superscript $(\overline{m})$. We use this notation to distinguish coefficients of a near boundary expansion (barred indices) from coefficients of a derivative expansion (unbarred indices). We find that 
\begin{equation}
	K_{\mu\nu}+\gamma_{\mu\nu}
	=
	\sum_{\overline{n}=1}^{D-1}
	\left(\frac{\overline{n}}{2}g_{\mu\nu}^{(\overline{n})}+\frac{1}{2}\left(2 S^{(\overline{n})}+k^{(\overline{n})}\right)\eta_{\mu\nu}+\mathcal{O}(\partial^{\overline{n}})\right)r^{2-\overline{n}}
	+\mathcal{O}(r^{2-D})\ ,
\end{equation}
where $\mathcal{O}(\partial^{\overline{n}})$ means terms of order $\overline{n}$ in a derivative expansion. Since the boundary energy momentum tensor is finite once the equations of motion are satisfied \cite{Balasubramanian:1999re,deHaro:2000xn}, then \eqref{E:KmnToTmn} guarantees that all terms of order $r^{\overline{n}}$ with $\overline{n} \leq D-2$ will vanish (note that this also implies $g_{MN}^{(\overline{n})}=\mathcal{O}(\partial^{\overline{n}})$ for $\overline{n} \leq D-2$.)
Thus,
\begin{equation}
\label{E:Kmunuv1}
	K_{\mu\nu}+\gamma_{\mu\nu}
	=
	\left(\frac{D-1}{2}g_{\mu\nu}^{(\overline{D-1})}+\frac{1}{2}\left(2 S^{(\overline{D-1})}+k^{(\overline{D-1})}\right)\eta_{\mu\nu}+\mathcal{O}(\partial^{D-1})\right)r^{3-D}
	+\mathcal{O}(r^{2-D})\ .
\end{equation}

Since we are interested only in second derivative corrections, we can neglect the $\mathcal{O}(\partial^{D-1})$ terms in \eqref{E:Kmunuv1} as long as $D>3$. We will treat the $D=3$ case separately in section \ref{S:AdS3}.
Using \eqref{E:Corrected} and $g^{\mu\nu}=\eta^{\mu\nu}+\mathcal{O}(r^{-1})$, the tracelessness condition \eqref{E:Traceless} reads
\begin{equation}
\label{E:Stoh}
	S^{(\overline{D-1})}=-\frac{(D-2)}{2}h^{(\overline{D-1})}\ .	
\end{equation}
Further, if we work in the Landau frame \eqref{E:Landau} then $j_{\lambda}^{(\overline{D-1})}$, $k^{(\overline{D-1})}$ and $h^{(\overline{D-1})}$ must satisfy
\begin{equation}
\label{E:LandauHolographic}
    j_{\lambda}^{(\overline{D-1})}P^{\lambda}_{\nu} =0\ ,\quad
    k^{(\overline{D-1})}+h^{(\overline{D-1})} = \frac{1}{b^{(D-1)}}\ .
\end{equation}
As explained in \cite{VanRaamsdonk:2008fp} for $D=4$ and in \cite{Bhattacharyya:2007jc} for $D=5$ and as we shall see more generally below, 
fixing these coefficients corresponds to choosing the zero momentum quasi-normal modes of the black hole. 
In other words, on the gauge theory side we fixed the ambiguity in determining the velocity field and temperature by going to the Landau frame.  On the black hole side, this ambiguity manifests itself in constant shifts of $u^\mu$ and $b$ which we fix by \eqref{E:LandauHolographic}.
We elaborate on this point in section \ref{S:Main}.
Using \eqref{E:Stoh} and  \eqref{E:LandauHolographic}, we find that in the Landau frame \eqref{E:KmnToTmn} reads
\begin{equation}
\label{E:EMTFormula}
	2 \kappa_D^{2} \langle T_{\mu\nu} \rangle = \frac{1}{b^{D-1}}\left((D-1)u_{\mu}u_{\nu}+\eta_{\mu\nu}\right)+(D-1)\pi^{(\overline{D-1})}_{\mu\nu}+\mathcal{O}(\partial^{D-1})\ .
\end{equation}
From \eqref{E:EMTFormula} we see that in order to obtain the hydrodynamic coefficients $\eta,\,\tau_{\Pi}$ and $\lambda_i$, we only need the ($D-1$)'th coefficient of $\pi_{\mu\nu}$ in a series expansion for large $r$. Since this coefficient will play an important role in what follows, we define
\begin{equation}
\label{BigPi}
    (D-1)\pi_{\mu\nu}^{(\overline{D-1})} \equiv \Pi_{\mu\nu}\ .
\end{equation}


\section{Solving the Einstein equations}
\label{S:Main}
To proceed, we need to solve \eqref{E:Einstein} order by order in a derivative expansion of the velocity fields. One method to keep track of derivatives in such an expansion is to solve the equations of motion in a neighborhood of some arbitrary point $x^{\mu}$, say $x^{\mu}=0$, and then extend the result to the whole manifold. See \cite{Bhattacharyya:2007jc} for further details. At the point $x^{\mu}=0$ we can of course set $u^{\mu}(0)=(1,\vec{0})$ and $b(0)=b_0$, which we do. Thus, after plugging the ansatz \eqref{E:Corrected} with the zero order solution \eqref{E:zeroorder} and expanding $S$, $k$, $h$, $j$ and $\pi$ around $x^{\mu}=0$, we find at order $n$
\begin{subequations}
\label{E:EOMD} 
\begin{align}
	\partial_r\left(r(1-b_0^{D-1}r^{D-1})\partial_r\pi^{(n)}_{ij}(r)\right) &= \mathbf{P}^{(n)}_{ij}(b_0 r)\\
	\partial_r \left(r^{D} \partial_r j^{(n)}_i(r)\right)&=\mathbf{J}^{(n)}_{i}(r)\\
\label{E:EOMS}
	(D-2) r^{-1}\partial_r S^{(n)}(r) - \frac{D-2}{2 r^2}\partial_r \left(r^2 \partial_r h^{(n)}(r)\right)&=\mathbf{S}^{(n)}(r)\\
\label{E:EOMK}
	\partial_r\left(r^{D-1} k^{(n)}(r)\right)+2 (D-1) r^{D-2} S^{(n)}(r)+
	\left(\frac{(D-3)}{2 b_0^{D-1}}-\frac{(D-2)}{r^{1-D}}\right)\partial_r h^{(n)}(r)
	&= \mathbf{K}^{(n)}(r)\,,
\end{align}
\end{subequations}
where $i=1,\ldots,D-2$. Note that since $\pi^{(n)}$ and $j^{(n)}$ are 
the order $n$ terms in a derivative expansion, 
only their spatial components appear in the expansion of the equations of 
motion around $x^{\mu}=0$ (because $u^\mu \pi_{\mu \nu} = 0$ and $u^\mu j_\mu=0$). The contributions coming from the known, order $n-1$, terms are contained in the sources on the right hand side of equation \eqref{E:EOMD}.
These have to be determined order by order.

In addition to \eqref{E:EOMD} there are $D$ first order constraint equations. Of these, $D-1$ are simply a statement of energy conservation of the boundary theory stress tensor
\begin{equation}
\label{E:Conservation}
 	\partial_{\mu}\langle T^{\mu\nu} \rangle = 0\ ,
\end{equation}
and restrict the relation between the temperature and velocity fields (at the appropriate order in a derivative expansion). For example, at second order in the derivative expansion, the derivative in \eqref{E:Conservation} acts on the first order energy momentum tensor. The resulting equations give the covariant version of the Navier Stokes equation. The $D$'th constraint equation should be related to tracelessness of the boundary theory stress tensor \cite{Gubser:2008vz}. Since we have already imposed tracelessness by hand via \eqref{E:Traceless} this is difficult to observe in the current formalism.

Recall that we have not completely fixed the gauge and we are still allowed to rescale the radial coordinate. One family of gauges used in \cite{Bhattacharyya:2007jc} and \cite{VanRaamsdonk:2008fp} which is consistent with \eqref{E:Stoh} is to set
\begin{equation}
	S = -\frac{(D-2)}{2}h\ .
\end{equation}
Another possibility discussed in \cite{Bhattacharyya:2008xc} is 
\begin{equation}
 	S=1\ .
\end{equation}
Here we shall use
\begin{equation}
\label{E:heq1}
 	h=1\ ,
\end{equation}
since then \eqref{E:EOMS} and \eqref{E:EOMK} reduce to first order equations. It is always possible to reach the gauge \eqref{E:heq1} by a redefinition of the radial coordinate, $\tilde{r}=r\sqrt{h}$.

After fixing the gauge \eqref{E:heq1}, the most general solution to \eqref{E:EOMD} contains $D(D-3)+2(D-2)+2$ integration constants: the second order equations for $\pi_{ij}$ and $j_i$ leave $D(D-3)$ and $2(D-2)$ undetermined coefficients, and the two first order equations for the scalars leave two more coefficients. Of these, $D(D-3)/2+(D-2)+1$ are fixed by requiring that there are no deformations of the asymptotically AdS boundary. This implies that the leading large $r$ behavior of all the fields should vanish. Requiring that the fields are smooth everywhere except at $r=0$ fixes the rest of the $D(D-3)/2$ integration constants for $\pi_{ij}$. The remaining $(D-2)+1$ integration constants are $j^{(\overline{D-1})}_i$ and $k^{(\overline{D-1})}$. These correspond to the solutions of the linearized Einstein equations at zero momentum, i.e., the zero momentum quasi-normal modes. Physically, they correspond to constant shifts in the temperature and velocity fields of the black hole, and we need a prescription for fixing their values. On the gauge theory side, such a redefinition of the velocity and temperature fields is also possible and we have fixed this ambiguity by going to the Landau frame \eqref{E:Landau}. As we have discussed in section \ref{S:BtB} the holographic counterpart of going to the Landau frame is \eqref{E:LandauHolographic}, which precisely fixes the zero momentum quasi normal modes of the black hole.
Thus, the solution to \eqref{E:EOMD} is
\begin{subequations}
\label{E:GenericSol}
\begin{align}
\label{E:piIntegral}
    \pi^{(n)}_{ij}(r) &= -\frac{1}{b_0}\int_{r}^{\infty}\frac{\int_1^x \mathbf{P}^{(n)}_{ij}(x^{\prime})dx^\prime}{x(1-x^{D-1})} dx \\
    j^{(n)}_i(r) &=-\int_r^{\infty} \frac{1}{x^D}\int^{x}_1 \mathbf{J}^{(n)}_i(x^{\prime})dx^{\prime} dx+r^{-(D-1)}C_i \\
    S^{(n)}(r) &=-\frac{1}{D-2}\int_r^{\infty} x \mathbf{S}^{(n)}(x) dx \\
    k^{(n)}(r) &=r^{-(D-1)}\int_1^{r}\left(\mathbf{K}^{(n)}(x)
    -2(D-1) x^{D-2} S^{(n)}(x)\right) dx + r^{-(D-1)} C_0\ ,
\end{align}
\end{subequations}
where $C_i$ and $C_0$ are chosen so that the $r^{-(D-1)}$ term in a near boundary series expansion of \eqref{E:GenericSol} vanishes.
Note that in order for the integrals to exist, i.e., in order to have an asymptotically AdS solution, we need that
\begin{equation}
\label{E:Existence}
	\mathbf{S} = \mathcal{O}(r^{-3})\ ,\quad
	\mathbf{K} = \mathcal{O}(r^{D-3})\ ,\quad
	\mathbf{P} = \mathcal{O}(r^{D-3})\ ,\quad
	\mathbf{J} = \mathcal{O}(r^{D-3})\ .
\end{equation}

What we are actually interested in is the $r^{-(D-1)}$ coefficient of a near boundary expansion of $\pi_{\mu\nu}(r)$. This is because \eqref{E:EMTFormula} implies that this coefficient is the only one that holds information about the boundary theory energy momentum tensor. 
After reinstating powers of $b$, expanding \eqref{E:piIntegral} in a power series, and using \eqref{BigPi}, we find
\begin{equation}
\label{E:Transport}
    b^D\Pi^{(n)}_{\mu\nu} =
        \lim_{\rbvar\to\infty}
            \left(
            \sum_{m=0}^{D-3}(-1)^{m+1}\frac{\rbvar^{m+1}\partial_\rbvar^m \mathbf{P}^{(n)}_{\mu\nu}(\rbvar)}
            {(m+1)!}+\int_1^\rbvar \mathbf{P}^{(n)}_{\mu\nu}(\rbvar')d\rbvar'
            \right).
\end{equation}
Perhaps it is worth emphasizing that when computing $\Pi^{(n)}_{\mu\nu}$, as long as the conditions in \eqref{E:Existence} are satisfied, we can completely neglect the source terms for the scalar and vector modes. This simplifies the analysis as compared to \cite{Bhattacharyya:2007jc} and \cite{VanRaamsdonk:2008fp}, where the full $D=5$ and $D=4$ dimensional black hole solutions were determined (to second order in a gradient expansion). In particular,
we are relieved of classifying all possible scalar and vector modes of $SO(D-2)$. In addition, we will see that $\mathbf{P}$ takes on a form which could be guessed from \eqref{E:Viscousfluid} so that we do not need to construct the traceless tensors of $SO(D-2)$ either.


\subsection{First order expansion}
Inserting \eqref{E:zeroorder} into \eqref{E:Corrected} and plugging it into \eqref{E:Einstein} we obtain the first derivative ($n=1$) terms in \eqref{E:GenericSol}. The source term $\mathbf{P}_{ij}$ is given by
\begin{equation}
\label{E:PSourceO1}
    b^{-2}\mathbf{P}^{(1)}_{\mu\nu}(r b)=
        (D-2)(rb)^{D-3} \sigma_{\mu\nu}\ 
\end{equation}
when restricted to the neighborhood of $x^{\mu}=0$.
The sources for the other modes obey the conditions in \eqref{E:Existence}.
From \eqref{E:Transport} we find
\begin{equation}
\label{E:Pi1}
   \Pi^{(1)}_{\mu\nu}=-\frac{1}{b^{D-2}}\sigma_{\mu\nu}
\end{equation}
implying that
\begin{equation}%
\label{E:EnergyMomentum1}
	2 \kappa_D^2 \langle T_{\mu\nu} \rangle = \frac{1}{b^{D-1}}\left((D-1)u_{\mu}u_{\nu}+\eta_{\mu\nu}\right)-\frac{1}{b^{D-2}}\sigma_{\mu\nu}\ .
\end{equation}
Using the relation \eqref{E:btoT}
together with \eqref{E:sdensity}
(or alternatively, by computing a quarter of the area of the horizon) we obtain \eqref{E:StoE}.

Had we been interested only in the first derivative corrections to the hydrodynamic energy momentum tensor we could have stopped here. Since we are looking for two derivative corrections, we need to have the explicit form of $S^{(1)}$, $j^{(1)}$, $k^{(1)}$ and $\pi^{(1)}$ since they will contribute later to $\mathbf{S}^{(2)}$, $\mathbf{J}^{(2)}$, $\mathbf{K}^{(2)}$ and $\mathbf{P}^{(2)}$. The various sources in \eqref{E:EOMD} are given by restricting
\begin{subequations}
\label{E:SourcesO1}
\begin{align}
    \mathbf{S}^{(1)}&=0\\
    \mathbf{K}^{(1)}&=2 r^{D-3}\partial_{\lambda}u^{\lambda}\\
    \mathbf{J}^{(1)}_{\nu}&=(D-2)r^{D-3}u^{\lambda}\partial_{\lambda}u_{\nu}
\end{align}
\end{subequations}
to $x^{\mu}=0$, together with \eqref{E:PSourceO1}. From \eqref{E:GenericSol} we find 
\begin{subequations}
\begin{align}
    S^{(1)}(r)&=0\\
    k^{(1)}(r)&=\frac{2}{D-2}\frac{1}{r} \partial_{\lambda}u^{\lambda}\\
    j^{(1)}_{\nu}(r)&=-\frac{1}{r}u^{\lambda}\partial_{\lambda}u_{\nu}\\
\notag
    \pi^{(1)}_{\mu\nu}(r)&=
	b \int_{b r}^\infty dx\,\frac{x^{D-2}-1}{x(x^{D-1}-1)}\sigma_{\mu\nu}\\
	&\equiv b F(b r)\sigma_{\mu\nu}\ .
\end{align}
\end{subequations}
The integral in the last line can be carried out explicitly,
\begin{equation}
	F(\rbvar) = 
	-\ln(\rbvar)
        +\sum_{n=1}^{D-2} \frac{\ln(\rbvar-r_n)r_n^{D-3}}{\sum_{k=0}^{D-3}(k+1)r_n^k}\ , \\
\end{equation}
where $r_n$ are the $D-2$ roots of the polynomial $\sum_{n=0}^{D-2} z^n = 0$. To facilitate the comparison with  
\cite{Bhattacharyya:2007jc,VanRaamsdonk:2008fp}
we can rewrite $F(\rbvar)$ in terms of arctangent functions and logarithms. This takes the somewhat bulky form
\begin{subequations}
\begin{align}
    F^{(D=4)}(\rbvar)=&-\ln(\rbvar)+\frac{1}{2}\ln(\rbvar^{2}+\rbvar+1)
-\frac{1}{\sqrt{3}} \arctan (\tfrac{2}{\sqrt{3}} \rbvar
+\tfrac{1}{\sqrt{3}})+\frac{\pi}{2\sqrt{3}}\ , \\
    F^{(D=5)}(\rbvar)=&-\ln(\rbvar)+\frac{1}{2}\ln(\rbvar+1)+\frac{1}{4}
\ln(1+\rbvar^{2})-\frac{1}{2}\arctan(\rbvar)+\frac{\pi}{4}\ , \\
    F^{(D=6)}(\rbvar)=&-\ln(\rbvar)- \frac{1}{2}\ln(2)
+\frac{1}{20}(5+\sqrt{5})\ln(2\rbvar^{2}+\sqrt{5}\rbvar+2+\rbvar)\\
&\mbox{} -\frac{1}{20}(-5+\sqrt{5})\ln(2\rbvar^{2}-\sqrt{5}\rbvar+2+\rbvar) \nonumber \\
&\mbox{} -\frac{1}{20}\sqrt{10+2\sqrt{5}}(\sqrt{5}-1)\arctan\Big(\frac{1}{40}
\sqrt{10-2\sqrt{5}}(5+\sqrt{5})(\sqrt{5}+4\rbvar+1)\Big) \nonumber \\
&\mbox{} +\frac{1}{20}\sqrt{10-2\sqrt{5}}(1+\sqrt{5})\arctan\Big(\frac{1}{40}
\sqrt{10+2\sqrt{5}}(-5+\sqrt{5})(-\sqrt{5}+4\rbvar+1)\Big) \nonumber \\
&\mbox{} +\frac{1}{40}\pi\sqrt{5}\sqrt{10+2\sqrt{5}}
+\frac{1}{40}\pi\sqrt{10+2\sqrt{5}}\ ,  \nonumber \\
    F^{(D=7)}(\rbvar)=&-\ln(\rbvar)+\frac{1}{12} \ln(1+\rbvar^{2}-\rbvar)
-\frac{1}{2\sqrt{3}} \arctan(\tfrac{2}{\sqrt{3}}\rbvar-\tfrac{1}{\sqrt{3}}) \\
&\mbox{} +\frac13\ln(\rbvar+1)+\frac14\ln(\rbvar^{2}+\rbvar+1)
-\frac{1}{2\sqrt{3}}\arctan(\tfrac{2}{\sqrt{3}}\rbvar+\tfrac{1}{\sqrt{3}})
+\frac{\pi}{2\sqrt{3}}\ . \nonumber
\end{align}
\end{subequations}

\subsection{Second order expansion}
At second order
the source $\mathbf{P}_{ij}$ is given by 
\begin{align}
\notag
 	b^{-3}\mathbf{P}_{\mu\nu}^{(2)}(\rbvar) =&
		\Big(2 \rbvar^{(D-2)/2}\partial_{\rbvar}\left(\rbvar^{(D-2)/2} F(\rbvar)\right)
			-(D-4)\rbvar^{D-4}\Big)\Sigma^{(0)}_{\mu\nu} \\
\notag
	&+\left( (D-2)\rbvar^{D-3}F(\rbvar)+\left(1-\rbvar^{D-1}\right)\rbvar (\partial_{\rbvar}F(\rbvar))^2
			-\frac12 (D-3)\rbvar^{D-4}\right) \Sigma^{(1)}_{\mu\nu} \\
\notag
	&+\Big(2 \rbvar^{D-4} + 4 \rbvar^{(D-2)/2}\partial_{\rbvar}\left(\rbvar^{(D-2)/2} F(\rbvar)\right)\Big) \Sigma^{(2)}_{\mu\nu} \\
\label{E:P2}
	&-\left( \frac{4}{\rbvar^3} + 2(D-3)\rbvar^{D-4} \right) \Sigma^{(3)}_{\mu\nu}
\end{align}
after expanding it around $x^{\mu}=0$. In \eqref{E:P2} we have used $\rho = b r$.
We remind the reader that $\Sigma^{(i)}=0$ for $D=3$ and $\Sigma^{(1)}=\Sigma^{(3)}=0$ for $D=4$.
One can now use \eqref{E:Transport} 
to compute $\Pi^{(2)}_{\mu\nu}$, and
\begin{equation}
	c = \frac{8\pi^2(d-1)^2b^{2-d}}{d \kappa_D^2} 
\end{equation}
to derive \eqref{E:Results}.
Of course, in order for a solution to exist at second order, we have to make sure that the sources satisfy \eqref{E:Existence}. They do.


\section{AdS${}_3$}
\label{S:AdS3}
AdS${}_3$ is dual to a $1+1$ dimensional conformal field theory. In $1+1$ dimensions there can be no transport coefficients since there is no non-trivial transverse symmetric traceless component of the energy momentum tensor. Thus, we expect that the energy momentum tensor will be that of a perfect fluid:
\begin{equation}
\label{E:2dfluid}
    \langle T_{\mu\nu}\rangle
    =\frac{\pi\mathfrak{c}}{6} T^2\left(2 u_{\mu}u_{\nu}+\eta_{\mu\nu}\right)\,,
\end{equation}
where $\mathfrak{c}$ is the central charge. To gain some more insight into conformal fluid dynamics in $1+1$ dimensions we switch to light-like coordinates
\begin{equation}
\label{E:lightlikecoords}
 	x^- = x-t\ , \quad x^+=x+t\,,
\end{equation}
where the velocity field takes the form:
\begin{equation}
\label{E:2dvelocity}
	u_{\mu}(x^-,x^+)dx^{\mu} = \frac{dx^-}{2\theta(x^-,x^+)}-\frac{1}{2}\theta(x^-,x^+)dx^+\,.
\end{equation}
In this coordinate system, the energy momentum tensor is given by
\begin{equation}
	\langle T_{\mu\nu} \rangle
	=
	\frac{\pi\mathfrak{c}}{12} T(x^-,x^+)^2
	\begin{pmatrix}
	\theta(x^-,x^+)^{-2} & 0 \\
	0 & \theta(x^-,x^+)^2
	\end{pmatrix}.
\end{equation}
Energy conservation,
\begin{equation}
\label{E:2dconstraint}
    \partial_{-}\langle T_{++}\rangle =0\quad
	\mbox{and}\quad
    \partial_{+}\langle T_{--}\rangle =0\,,
\end{equation}
implies that $T(x^-,x^+)$ and $\theta(x^-,x^+)$ should be decomposed such that
\begin{equation}
\label{E:bthetaToUV}
 	T(x^-,x^+)= \frac{1}{2\pi b_0}X_-(x^-) X_+(x^+)\ , \quad \theta(x^-,x^+) = \frac{X_-(x^-)}{X_+(x^+)}\ .
\end{equation}
Thus, we have traded the velocity and temperature dependence of the fluid with two holomorphic functions, $X_+$ and $X_-$.
Clearly, by going to a new coordinate system $\tilde{x}^-,\tilde{x}^+$ defined through
\begin{equation}
\label{E:rescaling}
	d\tilde{x}^- = X_-(x^-)^2dx^-\ , \quad
	d\tilde{x}^+ = X_+(x^+)^2dx^+\,,
\end{equation}
we can rescale the energy momentum tensor to be proportional to the identity: that of a fluid at rest with temperature $T=1/(2\pi b_0)$. Put differently, we can always find a coordinate system where the fluid seems static. In this case, all of the dynamics of the fluid will be captured by the coordinate transformation \eqref{E:rescaling}.

Since the $1+1$ dimensional conformal fluid has no transport coefficients, it is possible to guess the metric for the AdS${}_3$ dual description.
By direct computation one can show that the line element
\begin{equation}
\label{E:AdS3}
    ds^2 = -r^{2}\left(1-\frac{1}{r^{2} b^{2}}-\frac{2}{r}\partial_{\lambda}u^{\lambda}\right)u_{\mu}u_{\nu}dx^{\mu}dx^{\nu}
    +r^{2} P_{\mu\nu}dx^{\mu}dx^{\nu}
    -r u^{\lambda}\partial_{\lambda}\left(u_{\mu}u_{\nu}\right)dx^{\mu}dx^{\nu}
    -2 u_{\mu}dx^{\mu}dr
\end{equation}
is an exact solution to the Einstein equations as long as
\begin{equation}
	u^{\lambda}\partial_{\lambda} b = b \partial_{\lambda} u^{\lambda}\quad\mbox{and}\quad
	\partial_{\nu} b  = b  u^{\lambda} \partial_{\lambda} u_{\nu}\ .
\end{equation}
The boundary theory energy momentum tensor \eqref{E:KmnToTmn} takes the form \eqref{E:2dfluid} once we use
\cite{Brown:1986nw,Henningson:1998gx,Henningson:1998ey,Balasubramanian:1999re}
\begin{equation}
	\mathfrak{c} = \frac{12\pi}{\kappa_3^2}\ ,
\end{equation}
which is consistent with \eqref{E:p0c} for $d=2$.

Since the boundary theory energy momentum tensor can be brought to a diagonal form via a coordinate transformation, we expect that \eqref{E:AdS3} should also coincide with the standard non-rotating BTZ black hole after a similar coordinate transformation in the bulk. This is also expected on more general grounds \cite{Banados:1992gq,Carlip:1995qv}. Indeed, after using
\begin{equation}
\label{E:unboostedtoboosted}
	d\tilde{x}^+ = X_+(x^+)^2 dx^+\ , \quad d\tilde{x}^- = X_-(x^-)^{2} dx^-\ , \quad \tilde{r} = \frac{r}{X_-(x^-)X_+(x^+)}\,,
\end{equation}
where $X_+$ and $X_-$ are related to the velocity field and temperature through \eqref{E:bthetaToUV} and \eqref{E:btoT},
one obtains \eqref{E:AdS3} with $u^{\mu} = (-1,1)$, $b=b_0$ and $ds^2 = d\tilde{x}^+d\tilde{x}^-$. This is nothing but the static AdS${}_3$ black hole in the Eddington-Finkelstein coordinate system.

The solution \eqref{E:AdS3} is exact, we can also write it in the more familiar Fefferman-Graham coordinates,
\begin{equation}
\label{E:AdS3FG}
	ds^2 =
		-\frac{\left(z^2-4 b^2\right)^2}{16 b^4 z^2} u_{\mu}u_{\nu}dx^{\mu}dx^{\nu}
		+\frac{\left(z^2+4 b^2\right)^2}{16 b^4 z^2} P_{\mu\nu}dx^{\mu}dx^{\nu}
		+z^{-2} dz^2\ .
\end{equation}
This allows us to make contact with \cite{Kajantie:2007bn} which discusses a special case of \eqref{E:AdS3} (or \eqref{E:AdS3FG}) where the flow was boost invariant: the authors observed that if one works with the flat space line element
\begin{equation}
	ds^2 = -d\tau^2+\tau^2d\eta^2
\end{equation}
and the special solutions to the fluid equations
\begin{equation}
	u^{\mu} = (1,\,0)\ , \quad
	b = \frac{\tau}{\tau_0}
\end{equation}
with $\tau_0$ an integration constant,
then the AdS${}_3$ dual of this flow is given by \eqref{E:AdS3FG}. Of course, since \eqref{E:AdS3FG} is diffeomorphic to the BTZ black hole, it fits the general asymptotically AdS${}_3$ solutions described in \cite{Skenderis:1999nb}.


\section*{Acknowledgments}
We would like to thank S.~Bhattacharyya, G.~Cardoso, V.~Gra\ss, R.~Loganayagam and S.~Minwalla for useful discussions. 
This work is supported
in part by the European Community's Human Potential Program under
contract MRTN-CT-2004-005104 ``Constituents,
fundamental forces and symmetries of the universe'' and the Excellence Cluster
``The Origin and the Structure of the Universe'' in Munich.
M.~H.~is supported by the German
Research Foundation (DFG) within the Emmy-Noether-Program (grant numbers:
HA 3448/3-1 and HA 3448/5-1). A.Y.~is supported in part by the German Research Foundation and by the Minerva foundation.

\bibliographystyle{JHEP}
\bibliography{HigherO}

\end{document}